\documentclass[aip,graphicx,amsmath,amssymb,reprint]{revtex4-1}
\usepackage{graphicx}
\usepackage{dcolumn}
\usepackage{bm}
\usepackage[utf8]{inputenc}
\usepackage[T1]{fontenc}
\usepackage{mathptmx}
\usepackage{etoolbox}
\usepackage[]{units}
\usepackage{xcolor}

\DeclareUnicodeCharacter{0308}{\"}
\DeclareUnicodeCharacter{030C}{\v{ }}
\DeclareUnicodeCharacter{0301}{\'{ }}

\makeatletter
\def\@email#1#2{%
 \endgroup
 \patchcmd{\titleblock@produce}
  {\frontmatter@RRAPformat}
  {\frontmatter@RRAPformat{\produce@RRAP{*#1\href{mailto:#2}{#2}}}\frontmatter@RRAPformat}
  {}{}
}%
\makeatother
\draft 

\begin{document}


\title{Resonant excitation of  vortex gyrotropic mode via surface acoustic waves} 



\author{A. Koujok}
\affiliation{Fachbereich Physik and Landesforschungszentrum OPTIMAS, Rheinland-Pf\"alzische Technische Universit\"at Kaiserslautern-Landau, 67663 Kaiserslautern, Germany}
\author{A. Riveros} \affiliation {Escuela de Ingeniería, Universidad Central de Chile, 8330601 Santiago, Chile}
\author{D. R. Rodrigues} \affiliation {Department of Electrical and Information Engineering, Politecnico di Bari, 70126 Bari, Italy}
\author{G. Finocchio} \affiliation {Department of Mathematical and Computer Sciences, Physical Sciences and Earth Sciences, University of Messina, 98166, Messina, Italy}
\author{M. Weiler} \affiliation{Fachbereich Physik and Landesforschungszentrum OPTIMAS, Rheinland-Pf\"alzische Technische Universit\"at Kaiserslautern-Landau, 67663 Kaiserslautern, Germany}
\author{A. Hamadeh*} \affiliation{Fachbereich Physik and Landesforschungszentrum OPTIMAS, Rheinland-Pf\"alzische Technische Universit\"at Kaiserslautern-Landau, 67663 Kaiserslautern, Germany}
\email{hamadeh@rptu.de}
\author{P. Pirro} \affiliation{Fachbereich Physik and Landesforschungszentrum OPTIMAS, Rheinland-Pf\"alzische Technische Universit\"at Kaiserslautern-Landau, 67663 Kaiserslautern, Germany}

\date{\today}




\begin{abstract}
Finding new energy-efficient methods for exciting magnetization dynamics is one of the key challenges in magnonics. In this work, we present an approach to excite the gyrotropic dynamics of magnetic vortices through the phenomenon of inverse magnetostriction, also known as the Villari effect. We develop an analytical model based on the Thiele formalism that describes the gyrotropic motion of the vortex core including the energy contributions due to inverse magnetostriction. Based on this model, we predict excitations of the vortex core resonances by surface acoustic waves whose frequency is resonant with the frequency of the vortex core. We verify the model's prediction using micromagnetic simulations, and show the dependence of the vortex core's oscillation radius on the surface acoustic wave amplitude and the static bias field. Our study contributes to the advancement of energy-efficient magnetic excitations by relying on voltage-induced driven dynamics, which is an alternative to conventional current-induced excitations.
\end{abstract}


\maketitle 

Since their introduction, complementary metal-oxide-semiconductor (CMOS) based technologies have established themselves as indispensable tools on the pursuit of scientific advancement \cite{bota1970parameter,hosticka1979dynamic,popovic1986magnetotransistor,singh1989new,gary1994powerpc,burd1995energy,el2005cmos,el2009cmos}. High packing density, versatile scalability, and outstanding efficiency have enabled CMOS to shape modern computation. However, the continuous tendency towards miniaturization of electronics, while also needing to preserve or even increase efficiency, imposes limitations on CMOS technology \cite{isaac1998reaching,rairigh2005limits,haron2008cmos}. 
Complementing CMOS, spin wave based circuits have been nominated as promising candidates for data processing and non-conventional computing applications at the micro and nano scales\cite{kostylev2005spin,yu2013omnidirectional,vogt2014realization,pirro2021advances,chumak2021roadmap,wang2021magnonic,gartside2022reconfigurable,sun2022strain}. Magnonic circuits using only low-energy spin waves can be very efficient in terms of power consumption. Nevertheless, energy-efficient conversion from the magnonic to the electronic domain and vice versa remains an important challenge for magnonics. Various methods of spin-wave excitation have been proposed over the past years, perhaps the most common are the use of microwave antennas \cite{buttner2000linear,engebretson2003spatially,covington2002time,demokritov2006bose,an2013unidirectional},  spin-transfer torque (STT) \cite{slonczewski1996current,berger1996emission,kiselev2003microwave,madami2011direct} and spin-orbit torques \cite{collet2016generation,demidov2020spin,merbouche2022spin}. However, the energy efficiency of these techniques is still insufficient due to Ohmic losses, especially when scaling down to the nanoscale. Recently, voltage driven spin-wave excitation mechanisms have emerged \cite{bukharaev2018straintronics,sadovnikov2018magnon,li2021advances}. Utilizing electric fields, they are promising in terms of energy consumption \cite{bukharaev2018straintronics} since they efficiently minimize Joule heating \cite{mahmoud2020introduction}.

One way to realize a voltage driven spin-wave excitation which does not rely on the flow of electric currents is to couple spin waves to GHz surface acoustic waves (SAWs). SAWs are widely used for delaying and filtering of radio frequency signals. SAWs can be efficiently excited and detected using interdigital transducers (IDT) on a piezoelectric substrate via  inverse piezoelectricity or piezoelectricity, respectively \cite{tancrell1971wavefront,holland1974practical,maines1976surface,hashimoto2000surface,hess2002surface,weiler2011elastically,dreher2012surface,schmalz2020multi,ng2022excitation}. Applying a voltage at the electrodes of the IDT generates an electric field, and given that the IDT is fabricated onto a piezoelectric material, this electric excitation leads to the compression or expansion of the surface of the material, thus converting the electrical signal into mechanic displacement and strain. At the resonance frequency of the IDT, a propagating SAW is launched on the surface of the substrate.

One particularly interesting example of excitation of magnetization dynamics is found in magnetic vortices, whose dynamics have been widely investigated for various applications, such as non-volatile magnetic memories \cite{park2003magnetic,sousa2005non}, incorporation into existing spintronic devices such as spin-transfer torque oscillators \cite{hamadeh2014origin,locatelli2015efficient,hamadeh2023role}, and spin wave generation for electronics beyond the state of the art \cite{wintz2016magnetic,mayr2021spin,hamadeh2022hybrid}. On a more specific note, special interest has always been invested in the excitation and switching of the vortex core (VC) \cite{yamada2007electrical,guslienko2008dynamic,nakano2011all,ostler2015strain,hamadeh2023reversal2,iurchuk2023piezostrain,sun2023acoustic}, as it may be technologically exploited as a magnetic bit whereby data can be encoded as the core's polarization. Similar to magnetic skyrmions \cite{finocchio2016magnetic,woo2016observation}, the polarization of the VC is topologically stabilized. Its stability and exceptionally small size makes it a candidate as an information carrier.

In this work, we use propagating SAWs to drive a magnetic VC's gyrotropic motion via inverse magnetostriction. Based on the Thiele formalism, we phenomenologically model the gyrotropic dynamics of the VC in terms of its position in the plane of motion. After that, we utilize the GPU-accelerated micromagnetic simulation software Mumax3 \cite{ vansteenkiste2014design}, and the software platform Aithericon \cite{aithericon} to verify that the VC dynamics can be driven by longitudinal strain.

\begin{figure}[!ht]
  \includegraphics[width=\columnwidth]{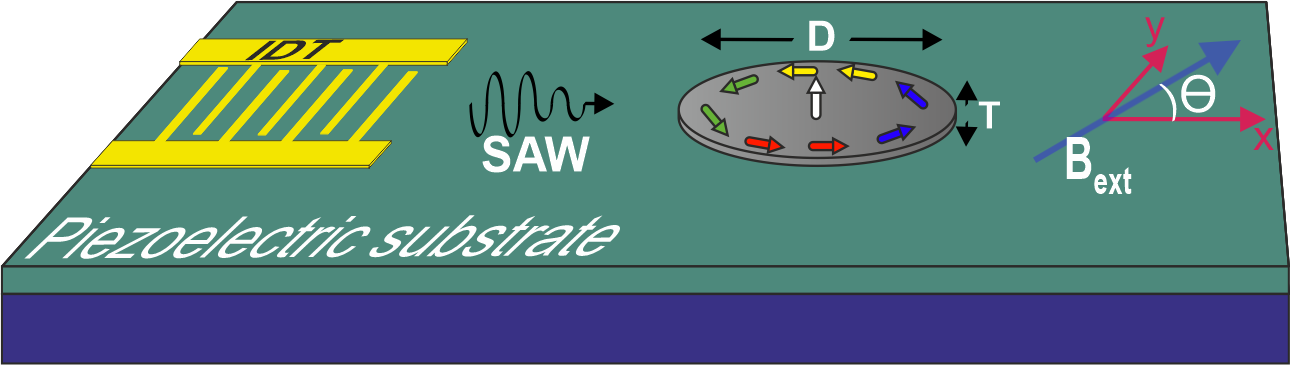}
  \caption{Schematic of the investigated system. The CoFeB disk, with a vortex magnetic configuration, has a thickness T = \unit[20]{nm} and a diameter D = \unit[500]{nm}. The disk is placed at a certain distance from the IDT which is used to excite SAWs. These waves propagate on the surface of the piezoelectric substrate towards the disk. The external magnetic field is applied in the film plane with an angle of $\Theta$ relative to the x-axis.}
  \label{fig:1}
\end{figure}

The investigated system (see FIG. \ref{fig:1}) consists of a Cobalt-Iron-Boron (CoFeB) disk in which the magnetic ground state is a vortex configuration. 
To ensure the stability of the magnetic vortex as the energy ground state of the disk, it is crucial to select a disk aspect ratio that ensures this stability. For this reason, the stability of the magnetic vortex state in magnetic dots has been widely investigated \cite{guslienko2001evolution,guslienko2008magnetic}. Here, we choose a disk having a thickness T = \unit[20]{nm} and a radius R = \unit[250]{nm}. The CoFeB disk is placed onto a piezoelectric material that allows electric-phononic conversion. Next to the CoFeB disk is an IDT to excite the SAW, which in turn propagates on the surface of the piezoelectric substrate and couples to the magnetization of the vortex. Here, we want to stress the fact that the SAW's wavelength is actually larger than the disk size, however it was presented in this manner so that the separation between the IDT's fingers coincides with half the SAW's wavelength.

Within the Thiele formalism of the Landau-Lifshitz-Gilbert equation, it is established that the VC's oscillation about its equilibrium position at the lowest frequency can be described as a function of the VC's position $\vec{R}$ \cite{guslienko2001field,guslienko2008magnetic,gaididei2010magnetic}. Employing the latter, we formulate the equation of motion describing the VC's gyrotropic motion, accounting for the different contributions to the effective field
\begin{equation}\label{eq:Thiele}
G[\dot{\vec{R}}\times\hat{z}] = \vec{F} - \eta\dot{\vec{R}}
\end{equation}
where $G = -2\pi pq$ is the gyroconstant depending on the vortex polarity $p$ and vorticity $q$. $\dot{\vec{R}}$ is the rate of change of the position of the vortex core in the x-y plane. $\eta \approx \pi\alpha\log(\frac{L}{l_\mathrm{ex}})$ is the viscosity constant dependant on the Gilbert damping $\alpha$, the disk radius $L$, and  the exchange length $l_\mathrm{ex}=\sqrt{2A/(\mu_{0}M_{s}^2)}$ where $A$ is the exchange stiffness, $\mu_{0}$ is the vacuum permeability, and $M_{s}$ is the saturation magnetization. $\vec{F}$ = -$\partial_{\vec{R}}E$ is the total force acting on the vortex, with $E$ being the total free energy of the vortex. Here, we assume an adiabatic evolution of the vortex, such that every position $\vec{R}$ of the core is associated to a single magnetization configuration, $\vec{m}$, and an associated total free energy, i.e. $\vec{m}\equiv\vec{m}(\vec{R})$ and $E\equiv E(\vec{R})$. The first term on the right hand side of Eq.~\eqref{eq:Thiele} is the precessional contribution and leads to the orbiting of the VC along equipotential curves of the energy landscape. The second term is the damping term, and leads the VC to move towards the minimum of the energy.

The applied strain generated by the SAW couples to the magnetization of the vortex via the Villari effect. It corresponds to a magneto-elastic energy term which is introduced as an additional contribution to the total free energy. Thus, to couple the VC to SAWs, we consider a total free energy given by $E(\vec{R}) = E_{0}(\vec{R}) + E_{SAW}(\vec{R})$, where the first term is the total energy of the magnetic vortex due to exchange, in-plane eaxy-axis anisotropy, Zeeman, and magnetostatic interactions, and the second term corresponds to the magnetoelastic coupling given by \cite{vanderveken2021finite}.

\begin{equation}
    E_{SAW} = T B_{1}\epsilon_{xx}\int  m_{x}^2 \, d^{2}x.
\end{equation}

Here, $T$ is the thickness of the disk and the integral is over the entire disk surface. We approximate the SAW strain profile and consider that the only non-vanishing magneto-elastic tensor component is the longitudinal strain component $\epsilon_{xx}$. Furthermore, $B_{1}$ is the first magneto-elastic coupling constant. The energy contributions were calculated analytically by considering the rigid vortex ansatz \cite{riveros2019surface,guslienko2001field}. Transversal strain components, namely $\epsilon_{yy}$ and $\epsilon_{zz}$, were neglected due to the assumption that the SAW was excited by means of a straight-line IDT. Specifically, the IDT excites Rayleigh wave modes where the primary displacements are predominantly in the vertical and longitudinal directions, with minimal horizontal motion. This ansatz correctly describes the vortex magnetization in an external magnetic field, although it could lead to some discrepancies as the VC approaches the disk edge. 
As an exemplary application of our model, we considered a CoFeB disk with a radius of \unit[250]{nm} and a thickness of \unit[20]{nm}. As a metallic magnetic material, CoFeB has a high saturation magnetization, which we set to $M_{s}$ = \unit[1150$\times$10$^{3}$]{A/m}. The exchange stiffness is \textbf{$A$} = \unit[15$\times$10$^{-12}$]{J/m}. Being a magnetostrictive material, the considered CoFeB disk also has a magnetoelastic coupling constant $B_{1}$ = \unit[-8$\times$10$^{6}$]{J/m$^3$}. Due to spin-orbit coupling the CoFeB disk exhibits uniaxial anisotropy, chosen to be along the x-axis with a strength given by $K$ = \unit[2900]{J/m$^3$}. These utilized magnetic parameters are conveyed from a previous experimental study of spin-waves driven by SAWs in CoFeB \cite{geilen2022parametric}. In the current study,  the effect of the SAW is considered by incorporating the magneto-elastic field generated by a spatially uniform strain oscillating at the frequency of the excited SAW. Due to the small disk size and the wave vector of the SAW being sufficiently small for frequencies resonant with the VC motion, this approximation is well justified. 
\begin{figure}[!ht]
  \includegraphics[width=\columnwidth]{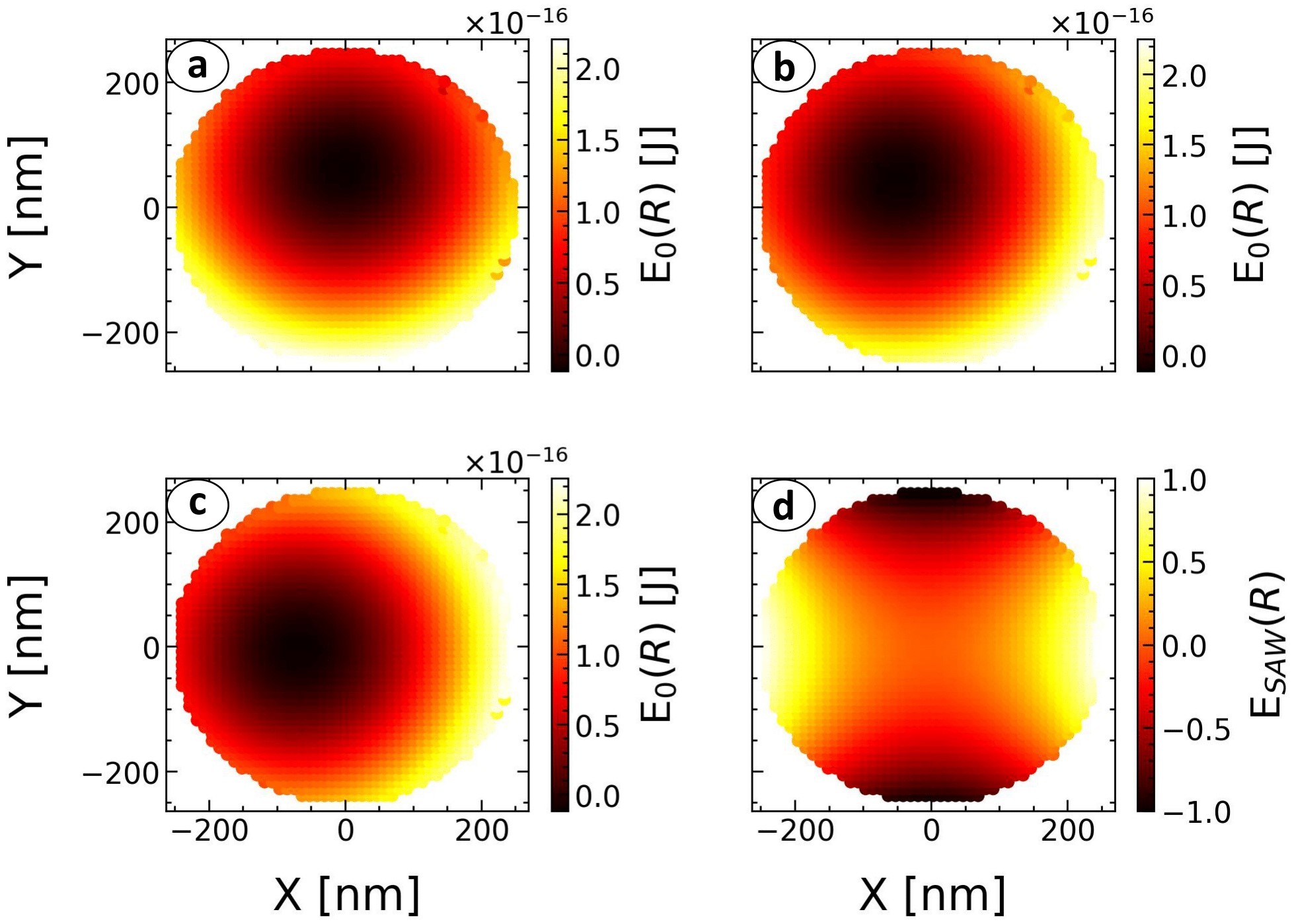}
 \caption{\textbf{Analytical Model:} (a,b,c) Energy landscape as a function of the VC position in the absence of strain at an in-plane field of \unit[20]{mT} applied (a) in the direction of the positive x axis, (b) at a 45 degree angle in the positive x-y directions and (c) in the direction of the positive y axis. (d) Landscape of the magneto-elastic energy as a function of the VC position.}
  \label{fig:2}
\end{figure}

FIG. \ref{fig:2}(a-c) show the total free energy landscape as a function of the VC position. If no magnetic field is applied, the energy is radially symmetric around the disk's center. For reasons explained below, we need to use asymmetric energy landscapes which can be achieved by applying in-plane fields. Then, the position corresponding to the minimal energy depends on the direction and strength of the applied magnetic field. To change the equilibrium position, we applied a magnetic field of \unit[20]{mT} in the (a) x-direction, (b) at an 45 degree angle in x-y, and (c) y-direction. FIG. \ref{fig:2}(d) represents the normalized magneto-elastic energy landscape. We notice that the center of the disk is a saddle-point of the magneto-elastic energy landscape, thus, at the center of the disk the total force due to the magnetoelastic coupling is zero. For this reason, to drive the VC with SAWs, it is necessary to move the VC stability position away from the disk center.

\begin{figure}[!ht]
  \includegraphics[width=\columnwidth]{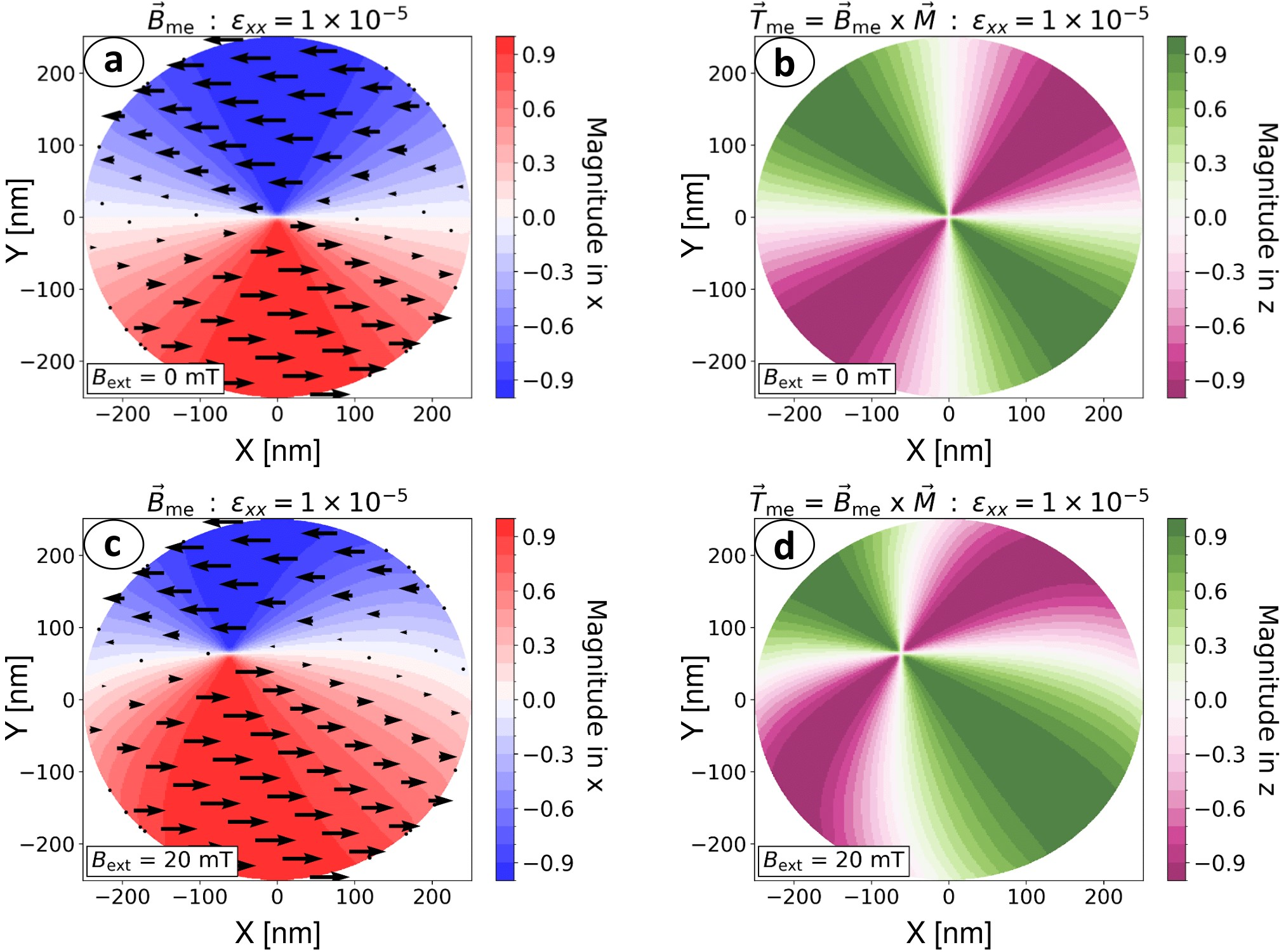}
 \caption{\textbf{Simulations:} (a,c) Magneto-elastic field distribution over the vortex at $\epsilon_{xx}$ = 1$\times$10$^{-5}$ in the absence of an applied magnetic field (a), and for $B_\mathrm{ext} = \unit[20]{mT}$ applied at a $\Theta$=45$^{o}$ angle with respect to the x-axis (c). The color code represents the normalized magneto-elastic field in the x-direction, while the vectors indicate their direction. (b,d) Magnetoelastic Torque profiles calculated as $\vec{B}_\mathrm{me}\times \vec{M}$ at $\epsilon_{xx}$ = 1$\times$10$^{-5}$ in the absence of an applied magnetic field (b), and for $B_\mathrm{ext} = \unit[20]{mT}$ applied at a 45$^{o}$ angle with respect to the x-axis (d).}
  \label{fig:3}
\end{figure}

The results of the Thiele model are compared to micromagnetic simulations using Mumax3 \cite{ vansteenkiste2014design} deployed onto Aithericon \cite{aithericon}. The simulations are performed for the disk geometry with vortex ground state presented in FIG. \ref{fig:1}. We incorporate the same material parameters, whereby a Landau-Lifshitz damping constant given by $\alpha$ = 0.004 is considered. To understand the  SAW driven VC gyration, it is useful to visualize the interaction of the magnetic vortex with external stimuli. In FIG. \ref{fig:3}, we demonstrate the importance of a symmetry breaking external applied in-plane field $B_{ext}$ for the VC  gyration by means of SAWs. We present the distribution profiles of the magnetoelastic field, which in this approximations, has only has a component  along the SAW propagation direction ($\vec{B}_{me}=-\frac{2}{M_s^2} B_1 \epsilon_{xx} \cdot M_x  \cdot \hat{e}_x $) , and its ensuing interaction with the magnetization, namely the magnetoelastic torque $\vec{T}_\mathrm{me}$ calculated as $\vec{T}_\mathrm{me}$ = $\vec{B}_\mathrm{me}\times \vec{M}$. Mumax3 adds $\vec{B}_\mathrm{me}$ as an additional contribution to the effective field $\vec{B}_\mathrm{eff}$ and solves the Landau-Lifshitz Gilbert equation. For $\epsilon_{xx}$ = 1$\times$10$^{-5}$ in the absence of an applied in-plane field (see FIG. \ref{fig:3} (a, b)), $\vec{B}_\mathrm{me}$ and $\vec{T}_\mathrm{me}$ are symmetrically distributed across the vortex structure. The equal yet opposite distribution of the magnetoelastic torque across the magnetic landscape leads to the cancellation of the desired driven gyration of the VC under the influence of the magnetoelastic torque. To break the spatial symmetry, we apply an in-plane magnetic field of \unit[20]{mT} strength at a 45$^{o}$ angle with respect to the x-axis (see FIG. \ref{fig:3} (c, d)). This implies a variance in the distribution of $\vec{B}_\mathrm{me}$ and $\vec{T}_\mathrm{me}$ across opposing parts in the magnetic landscape.The rather intriguing perpendicular displacement of the VC in response to the applied field is a direct consequence of the system's attempt to minimize the Zeeman energy of the in-plane curling domain. When a field is applied in the plane of the vortex, the in-plane domain tends to align to in the direction of the bias field. This leads to the expansion of the domain, instigating a displacement of the vortex core in a direction perpendicular to the field itself.

\begin{figure}[!ht]
  \includegraphics[width=\columnwidth]{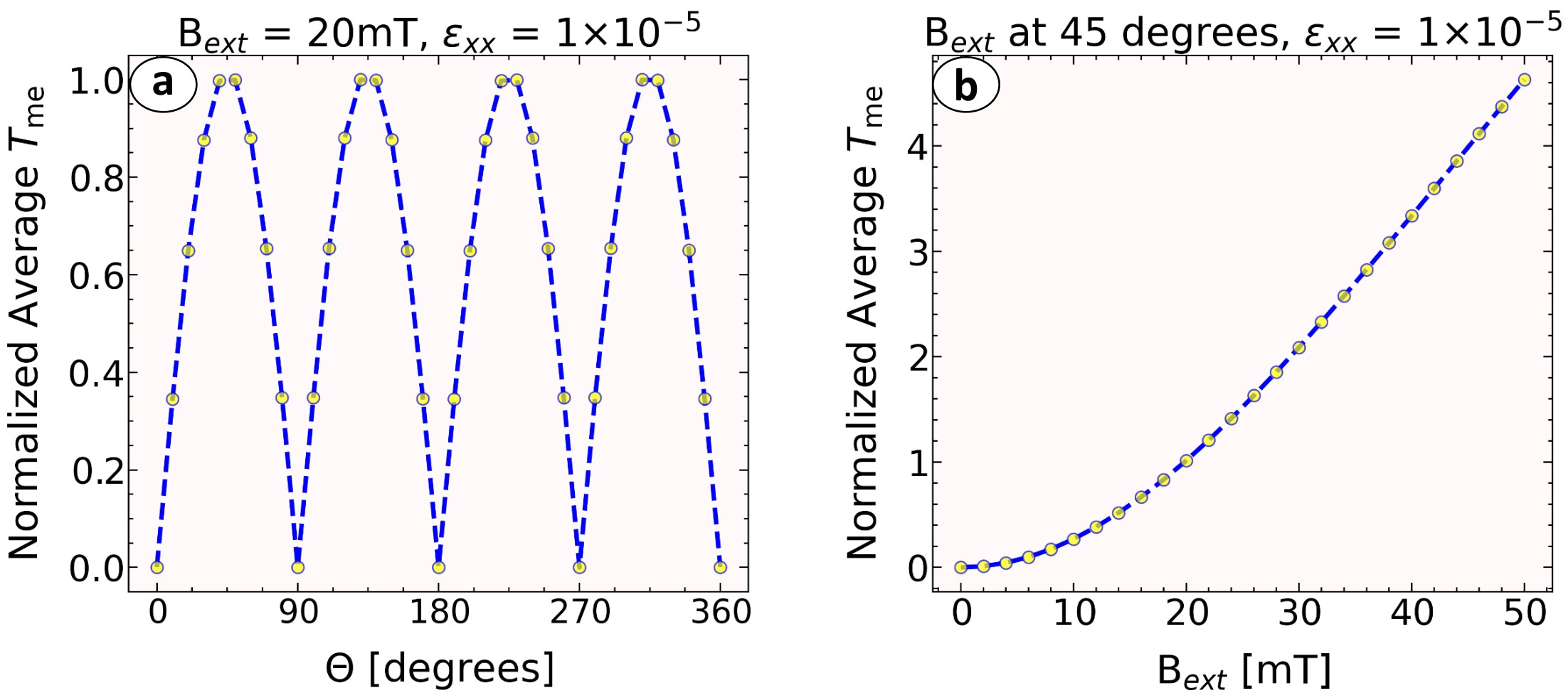}
 \caption{\textbf{Simulations:} (a) The variation of the average magneto-elastic torque as a function of the magnetic field's angle $\Theta$ at fixed field and strain values of \unit[20]{mT} and $\epsilon_{xx}$ = 1$\times$10$^{-5}$ respectively. (b) The variation of the average magneto-elastic torque as a function the magnetic field's magnitude for fixed $\Theta$ and strain values of 45$^{o}$ and $\epsilon_{xx}$ = 1$\times$10$^{-5}$ respectively.}
  \label{fig:4}
\end{figure}

To adopt an efficient mechanism of the VC's driven gyration, we study the effect of varying $\Theta$, the angle at which the magnetic field is applied, on the average magneto-elastic torque $T_\mathrm{me}$ (see FIG. \ref{fig:4} (a)). The external magnetic field's magnitude is fixed at \unit[20]{mT}, and that of the strain at $\epsilon_{xx}$ = 1$\times$10$^{-5}$. To calculate the magnitude of $T_\mathrm{me}$, we first evaluate $\vec{T}_\mathrm{me}$ = $\vec{B}_\mathrm{me}\times \vec{M}$ for the three magnetization components, then integrate over the entire vortex structure. We then calculate the modulus encompassing the three components of the torque vector, then average over the number of cells. As can be seen, the magneto-elastic torque is minimal for magnetic field  applied at 0$^{o}$, 90$^{o}$, 180$^{o}$, 270$^{o}$ with respect to the x-axis \cite{geilen2022fully,geilen2022parametric}. Therefore to enhance the magnetoelastic effect we choose the applied magnetic field at an angle of maximum torque, for instance at 45$^{o}$ and we keep this angle for the rest of our simulations. Furthermore, we study the change in torque in regard to the magnitude of the applied field, whilst also fixing the strain at $\epsilon_{xx}$ = 1$\times$10$^{-5}$ (see FIG. \ref{fig:4} (b)). Ranging between \unit[0]{mT} and \unit[50]{mT}, the torque steadily grows with increasing applied magnetic field, where the VC approaches the edge at \unit[50]{mT}. Consequently, we adopt an intermediate magnetic field value of \unit[20]{mT} for further simulations.

\begin{figure}[!ht]
  \includegraphics[width=\columnwidth]{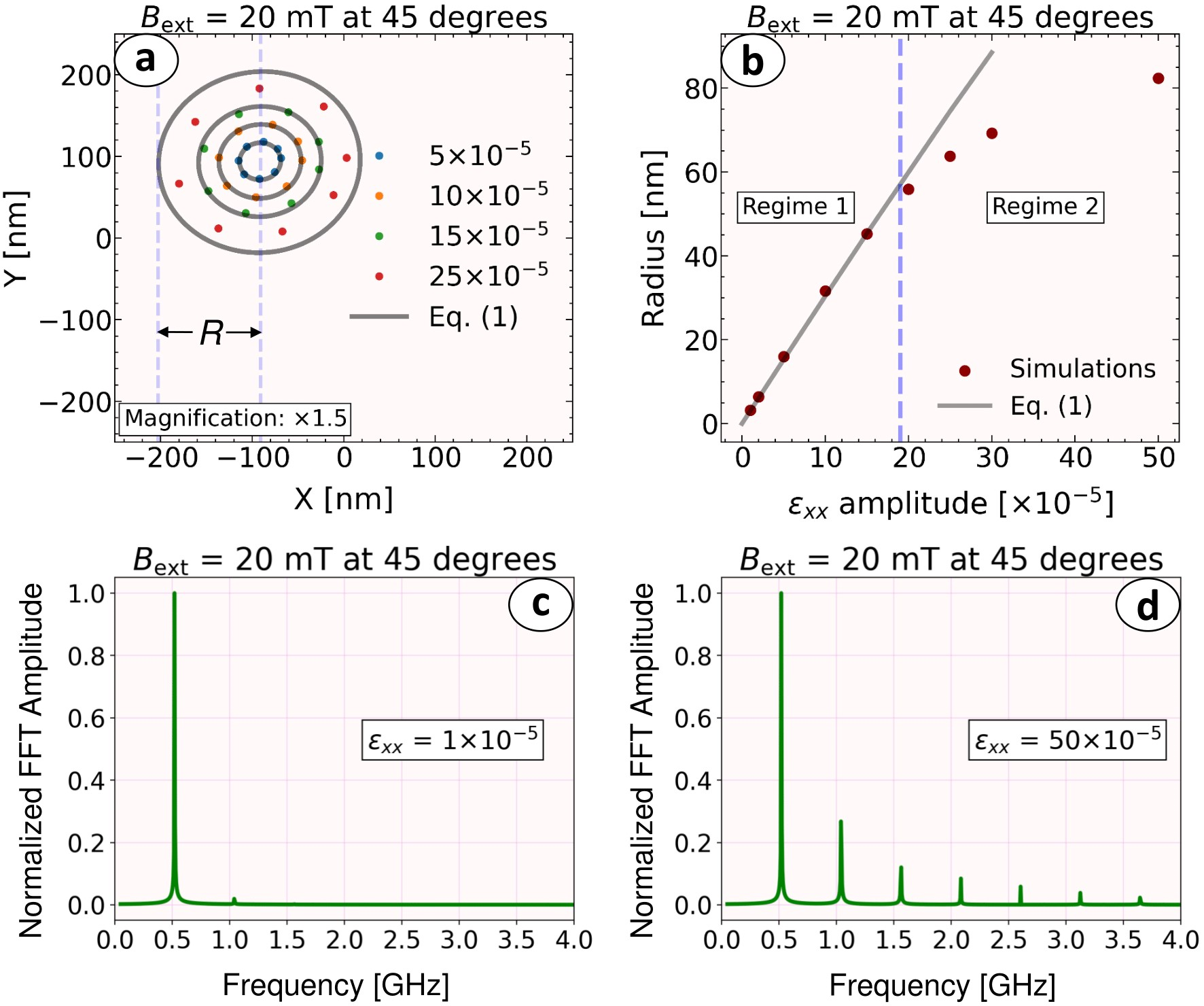}
 \caption{(a) Trajectories underwent by the VC for an applied longitudinal strain $\epsilon_{xx}$ at the gyrotropic mode's frequency by simulations (dots) and the analytical model (black lines). Presented $\epsilon_{xx}$ amplitudes are 5$\times$10$^{-5}$, 10$\times$10$^{-5}$, 15$\times$10$^{-5}$ and 25$\times$10$^{-5}$. $R$ is the radius of each trajectory of the driven VC, however it is highlighted only for the case of 25$\times$10$^{-5}$ for clarification purposes. (b) The variation of the trajectories' radii as a function of the strain's amplitude as per micromagnetic simulations (dots), and the fitting (black line) from the model (see Eq.~\eqref{eq:Thiele}). (c,d) FFT spectra of the simulated temporal magnetization components $M_{x}+iM_{y}$ for two cases at which the strain rf signal has an amplitude (c) 1$\times$10$^{-5}$ and (d) 50$\times$10$^{-5}$.}
  \label{fig:5}
\end{figure}

To investigate the SAW-driven dynamics of the VC, we first extract the gyrotropic frequency of the displaced VC. We emphasize that this frequency depends on the direction and amplitude of the applied constant magnetic field. We generated a short displacement of the vortex by an in-plane magnetic field, and performed the Fourier analysis of the resulting motion. For the adopted case of an applied static field of \unit[20]{mT} at a 45$^{o}$ angle with respect to the x-axis, the obtained gyrotropic mode's frequency is \unit[521]{MHz}. We apply a rf strain signal corresponding to the tensorial longitudinal strain component $\epsilon_{xx}$ sinusoidally oscillating at the VC's gyrotropic frequency (\unit[521]{MHz}). The excited SAW wavelength is considerably longer than the disk diameter, which based on this previous experimental study that investigates Rayleigh waves' excitation \cite{geilen2020interference}, is around $\unit[8.22]{\mu m}$. This results in a strain which is spatially uniform over the vortex structure. $\epsilon_{xx}$ is applied at different amplitudes: 1$\times$10$^{-5}$, 10$\times$10$^{-5}$, 15$\times$10$^{-5}$ and 25$\times$10$^{-5}$ as can be seen in FIG. \ref{fig:5} (a). FIG. \ref{fig:5} (a) shows that the VC underwent gyrotropic motion orbiting its equilibrium position defined by the external field in response to the driving force. Both micromagnetic and analytical results are shown, where the dotted orbits correspond to data extracted from the simulations, and the lined orbits to data from the model. We notice the inflation of the orbits as the amplitude of $\epsilon_{xx}$ is increased from 5$\times$10$^{-5}$ to 25$\times$10$^{-5}$. Given that the gyrotropic motion of the VC is driven at a constant frequency, namely the frequency of the SAW, an increase in the radius of the orbit does not necessarily affect the angular velocity. The latter is constant throughout, and is independent of the VC's position. The system compensates the growth of the orbit's radius by an increase in the linear velocity of the gyrating VC. This compensation ensures that the angular frequency only depends on the frequency of the oscillation, which in this case is equal to \unit[521]{MHz}. For better elucidation, refer to the "Supplementary Material" section. We present two videos showing the gyrotropic motion of the VC for $\epsilon_{xx}$ = 10$\times$10$^{-5}$ and $\epsilon_{xx}$ = 25$\times$10$^{-5}$. Both videos are produced for the last \unit[5]{ns} of the simulation.

It is noticeable that the model's fitting to the micromagnetic simulations starts to falter at large strain amplitudes (see $\epsilon_{xx}$ = 25$\times$10$^{-5}$). As a result, FIG. \ref{fig:5} (b) serves as a demonstration of the induced growth in the radius of the VC's trajectory in response to the strain amplitude. Herein, the term ”radius” refers specifically to the radius of the vortex motion, and not to the radius of the VC itself. In other words, we are referring to the radius of the trajectory of the VC under the influence of SAWs. In the linear regime (regime 1), that is at low strain values between 1$\times$10$^{-5}$ and 20$\times$10$^{-5}$, the analytical evaluations and the simulation extracted data of the trajectory's radius are in complete agreement. As the strain amplitude is increased beyond 20$\times$10$^{-5}$, this fitting starts to diverge (regime 2) as inferred by the non-linear growth of the simulated radius in comparison to the model. This is expected since the analytical model does not take into account the deformations of the vortex structure, which are non-local and highly dependent on the boundary conditions. This leads to a shift of the stability position, and to a different growth of the gyration radius in the limits of large strain amplitudes. To further elaborate on the difference of the simulations from the model, we present the Fast Fourier Transformation of the temporal magnetization components $M_{x}+iM_{y}$ for $B_{ext}$ = \unit[20]{mT} applied at 45$^{o}$ (see FIG. \ref{fig:5} (c, d)). In the linear regime at $\epsilon_{xx}$ = 1$\times$10$^{-5}$ (see FIG. \ref{fig:5} (c)), the FFT spectrum conveys that of a magnetic vortex with gyrotropic modal excitation. In contrast, different modal excitations appear at higher strain amplitudes inferred by the appearance of various peaks at higher harmonics of the gyrotropic mode's frequency in FIG. \ref{fig:5} (d). As non-linear effects become more prominent, the growth of the radius of the VC's trajectory can no longer be precisely described within the proposed model.

In conclusion, via micromagnetic simulations and analytical model calculations, we have demonstrated the feasibility of driving the gyrotropic motion of a vortex core (VC) using surface acoustic waves (SAWs). The proposed device uses the longitudinal strain tensorial component $\epsilon_{xx}$ of the SAW where the vortex gyration is caused by the Villari effect. The presented model is based on the Thiele formalism of the Landau-Lifshitz-Gilbert (LLG) equation which allows for quantitative evaluation of the system energies, whilst the simulations solve the LLG equation for vortex magnetization dynamics incorporating the strain contribution as the magnetoelastic field. The magnetic disk uses standard CoFeB material parameters and the required SAW frequency and magnitude can be experimentally realized \cite{kavalerov2000observation,shah2023symmetry}.
The ability to control the dynamics of the VC using energy-efficient manipulation methods, such as the one presented in this study, holds significant importance in the field of engineering nanosized magnetic vortices for permanent data storage applications. Additionally, optimizing energy conservation is a crucial aspect of ongoing research in this area. Overall, our findings contribute to the advancement of knowledge and pave the way for future investigations and developments in the field of magnetic vortex-based devices.

\section*{Supplementary Material}
For a demonstration of the vortex core's velocity dependence on the radius of the gyrotropic orbit, please refer to the supplementary material.

\section*{Acknowledgements}
This work has been  supported  by the European Research Council within the Starting Grant No. 101042439 "CoSpiN", Consolidator Grant No. 101044526 "MAWiCS" and by the Deutsche Forschungsgemeinschaft (DFG, German Research Foundation) - TRR 173 - 268565370" (project B01). AR has been supported by CIP2022036. DRR has been supported by the Italian Ministry of University and Research (MUR) within the D.M. 10/08/2021 n. 1062 (PON Ricerca e Innovazione). DRR and GR have been supported by the Italian Ministry of University and Research (MUR) within the project PRIN 2020LWPKH7 and by PETASPIN association (www.petaspin.com). We would like to extend our gratitude to Kei Yamamoto from Advanced Science Research Center, Japan Atomic Energy Agency,  Tokai , Japan for his insightful discussions.

\section*{References}

\bibliography{references}

\end{document}